\documentclass[sigconf]{acmart}
\AtBeginDocument{%
  \providecommand\BibTeX{{%
    \normalfont B\kern-0.5em{\scshape i\kern-0.25em b}\kern-0.8em\TeX}}}

\copyrightyear{2023}
\acmYear{2023}
\setcopyright{acmlicensed}
\acmConference[MM '23]{Proceedings of the 31st ACM International Conference on Multimedia}{October 29-November 3, 2023}{Ottawa, ON, Canada}
\acmBooktitle{Proceedings of the 31st ACM International Conference on Multimedia (MM '23), October 29-November 3, 2023, Ottawa, ON, Canada}
\acmPrice{15.00}
\acmDOI{10.1145/3581783.3611886}
\acmISBN{979-8-4007-0108-5/23/10}

\usepackage{multirow}
\usepackage{bbding}
\usepackage{subfig}
\usepackage{bm}

\usepackage{dsfont}
\usepackage{threeparttable}
\usepackage{amsmath}
\usepackage{amsfonts}
\usepackage{color}
\usepackage{comment}
\usepackage{graphicx}
\usepackage{multirow}

\usepackage{url}
\usepackage{booktabs}  
\usepackage{enumitem}
\usepackage{float}
\usepackage{array}
\usepackage{float}

\usepackage{algorithm}
\usepackage{algorithmic}
\usepackage[switch]{lineno}

\begin{document}
\title{Semantic-Guided Feature Distillation for Multimodal Recommendation}
\author{Fan Liu}
\affiliation{%
  \institution{National University of Singapore}
  \country{}
  }
\email{liufancs@gmail.com}

\author{Huilin Chen}
\affiliation{%
  \institution{Hefei University of Technology}
  \country{}
  }
\email{ClownClumsy@outlook.com}

\author{Zhiyong Cheng}
\affiliation{%
  \institution{Qilu University of Technology (Shandong Academy of Sciences)}
  \country{}
  }
\email{jason.zy.cheng@gmail.com}

\author{Liqiang Nie}
\affiliation{%
  \institution{Harbin Institute of Technology, Shenzhen}
  \country{}
  }
\email{nieliqiang@gmail.com}

\author{Mohan Kankanhalli}
\affiliation{%
  \institution{National University of Singapore}
  \country{}
  }
\email{mohan@comp.nus.edu.sg}


\begin{abstract}
Multimodal recommendation exploits the rich multimodal information associated with users or items to enhance the representation learning for better performance. In these methods, end-to-end feature extractors (e.g., shallow/deep neural networks) are often adopted to tailor the generic multimodal features that are extracted from raw data by pre-trained models for recommendation. However, compact extractors, such as shallow neural networks, may find it challenging to extract effective information from complex and high-dimensional generic modality features. Conversely, DNN-based extractors may encounter the data sparsity problem in recommendation. To address this problem, we propose a novel model-agnostic approach called Semantic-guided Feature Distillation (SGFD), which employs a teacher-student framework to extract feature for multimodal recommendation. The teacher model first extracts rich modality features from the generic modality feature by considering both the semantic information of items and the complementary information of multiple modalities. SGFD then utilizes response-based and feature-based distillation loss to effectively transfer the knowledge encoded in the teacher model to the student model. To evaluate the effectiveness of our SGFD, we integrate SGFD into three backbone multimodal recommendation models. Extensive experiments on three public real-world datasets demonstrate that SGFD-enhanced models can achieve substantial improvement over their counterparts. 
\end{abstract}




\maketitle

\section{Introduction}

Recommendation system plays an essential role in various online platforms, such as E-commerce, social media, and streaming services. Many recommendation approaches are based on Collaborative Filtering (CF)~\cite{Koren2009MF,netflix,he2017neural,He@LightGCN,Liu2021IMP_GCN,wang2022user,liu2023hs,Cheng2023CGCN}, which efficiently leverages user-item interactions. 
Nevertheless, the CF-based recommendation model is incapable of accurately capturing user preferences when the interaction data are sparse, resulting in poor recommendation performance. To tackle the limitation, many methods have been developed to leverage the side information, such as attributes, reviews, and images, associated with users and items, which provides rich information about user preference and item characteristics~\cite{mcauley2015image,he2016vbpr,chen2018attention,Liu2020A2GCN,Zhao2023CoGCN}. Recognizing that features from different modalities can be complementary to each other, various approaches have been proposed to exploit multimodal features for learning more comprehensive user and item representations for recommendation~\cite{zhang2017joint,hsieh2017cml,liu2018MAML,wei2019mm,Wei2019GRCN,Liu2022DMRL,BM32020Arxiv,wang2023generative,Wei2023LightGT}. 

\begin{figure*}[t]
	\centering
	\includegraphics[width=0.9\linewidth]{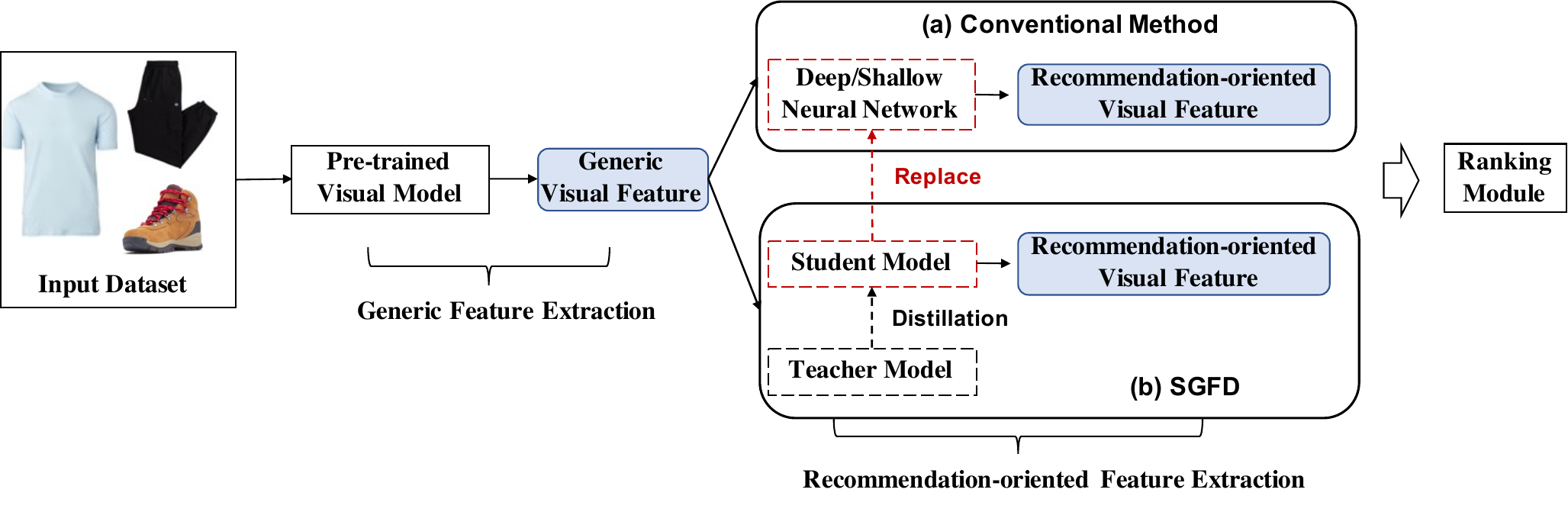}
	\caption{An illustration of the feature extraction pipeline in multimodal recommendation models. (a) is the conventional feature extraction method and (b) is our proposed SGFD.}
	\vspace{0pt}
	\label{fig:mecari}
\end{figure*}

Extracting effective features from multimodal data serves as a foundation for a multimodal recommendation model. Fig.~\ref{fig:mecari} uses the visual feature extraction process to illustrate the general pipeline for feature extraction in the multimodal recommendation methods. Typically, pre-trained models (such as VGG~\cite{Simonyan2015VeryDC} for visual features and Bert~\cite{Devlin2019BERTPO} for textual features) are firstly adopted to extract generic features, which are then tailored for recommendation via a designed feature extractor (e.g., shallow/deep neural network).
The multimodal features distilled from the feature extractor are subsequently used as input in the recommendation model to learn user and item representations. 
For example, Liu et al.~\cite{liu2018MAML} presented a metric learning-based method that models the diverse preferences of users by exploiting the multimodal features of the item to estimate the special attention of the user to different aspects of that item. To prune potential false-positive edges in the user-item interaction graph, Wei et al.~\cite{Wei2019GRCN} designed a method that scores the affinity between user preference and multimodal features of an item to measure the confidence of edge being true-positive interaction. It is worth mentioning that, in those methods, the feature extractor is often jointly learned with the user and item representations in the recommendation model in an end-to-end manner. 


Despite the progress, many existing methods overlooked the importance of feature extraction in multimodal recommendation.
Due to the complexity and high dimensionality of multi-modal features, a compact feature extractor adopting shallow neural networks (SNNs) can hardly extract effective information from the generic features of each modality. Therefore, most recent multimodal recommendation models employ deep neural networks as the extractor~\cite{zhang2017joint,hsieh2017cml,liu2018MAML,wei2019mm,Wei2019GRCN}, leveraging the powerful representation learning capability of DNNs. However, the problem is that deep models need sufficient training data for good performance. Unfortunately, recommendation models often face serious data sparsity problem. As a result, feature extractors with deep models cannot be well trained, which not only renders them unhelpful but can also negatively impact the final recommendation performance. Therefore, the primary challenge faced by existing multimodal recommendation methods is the dilemma of choosing between deep or shallow models as feature extractors, which is an essential part of the multimodal recommendation model.
The second challenge is that semantic information is often ignored during the feature extraction process. In fact, existing recommendation models extract modality features only using collaborative information~\cite{liu2018MAML,Wei2019GRCN,BM32020Arxiv}. Yet, semantic information contains rich semantic and concise knowledge, which can well guide the feature extraction process. For example, category information is a more abstract and general form of semantic information, and it can help recommendation models identify relationships or similarities between items.

Inspired by the above considerations, in this paper, we introduce a novel model-agnostic approach, termed \textbf{S}emantic-\textbf{G}uided \textbf{F}eature \textbf{D}istillation (SGFD for short), which can robustly extract effective recommendation-oriented features from generic multimodal features using the teacher-student framework.
Specifically, we propose a teacher model composed of feature extractors with deep neural networks for different modalities. These feature extractors are designed to extract features considering both the semantic information of items and the complementary information of multiple modalities. As the teacher model is separately trained with recommendation model, it will not suffer the data sparsity problem in recommendation. Meanwhile, we use the feature extractors with shallow neural networks as the student model to replace the original feature extractors in recommendation methods. 
In addition, we introduce a feature-based distillation loss and a response-based distillation loss to transfer knowledge from the teacher model to the student model. 
To evaluate the efficacy of our approach, we conducted comprehensive experiments on three real-world datasets. The results demonstrate that our method can significantly outperform existing multimodal recommendation models. We have released the code and relevant parameter
settings to facilitate repeatability as well as further research~\footnote{https://github.com/HuilinChenJN/SGFD.}.

In summary, the main contributions of this work are as follows:
\begin{itemize}
\item We emphasize the constraints of existing recommendation methods in effectively utilizing multimodal features. To overcome these limitations, we propose a model-agnostic approach SGFD that can extract effective recommendation-oriented features from generic modality features using the teacher-student framework.

\item To leverage the semantic information during the feature extraction processes, we design feature extractors that consider both the semantic information of items and the complementary information of multiple modalities. 

\item We employ the response-based and feature-based distillation loss to facilitate knowledge transfer from the teacher to the student model. 

\item We have conducted extensive experiments on three real-world datasets, demonstrating the superiority of our method.
\end{itemize}


\section{related work}
\label{sec:relatedwork}

\subsection{Multimodal Recommendation}
Various types of side information have been utilized in recommender systems, such as attributes~\cite{chen2018attention,Liu2020A2GCN}, reviews~\cite{catherine2017transnets,mcauley2013hidden,tan2016rating,cheng20183ncf,chen2018neural,cheng2018aspect,chin2018anr} and images~\cite{kalantidis2013getting,wang2017your,he2016vbpr,mcauley2015image}. Most existing models exploit different types of information individually rather than fusing them, due to the challenges involved in effectively combining different modalities of information. However, it is well-recognized that multimodal features can provide complementary information to each other, as demonstrated in~\cite{zhang2016collaborative,zhang2017joint,liu2018MAML,SunWSFN22,sun2022counterfactual,Wang2023MIRec}. Thus, various approaches have been proposed to learn the user and item representations by fusing multimodal features. For example, JRL~\cite{zhang2017joint} first extracts user and item features from ratings, reviews, and images separately with deep neural networks, then concatenates those features to form the final user and item representations. 
In recent years, Graph Convolution Networks (GCNs) have attracted increasing attention in multimedia recommendation due to their powerful capability on representation learning from non-Euclidean structure~\cite{wei2019mm,Wei2019GRCN,Zhang2021MLS}. MMGCN~\cite{wei2019mm} learns the model-specific user preference to the content information via the direct information interchange between user and item in each modality. 
For reducing the computational cost on large graphs and removing noise supervision signals caused by the negative sampling strategy, BM3~\cite{BM32020Arxiv} introduces a self-supervised method which can learn the representations of users and items by jointly optimizing three multimodal objectives.

Despite the progress, the feature extractor in these methods is jointly optimized with the CF-based recommendation approach. However, a compact extractor may not extract effective modality features, and a DNN-based extractor suffers from data sparsity in recommendation.
In this paper, to utilize more useful multimodal information for user and item representation learning, we employ the knowledge distillation technique to extract more effective recommendation-orient multimodal features from generic multimodal features.

\subsection{Knowledge Distillation}
Knowledge Distillation (KD) is composed of a “compact” model (i.e., student model) and a previously trained “large” model (i.e., teacher model)~\cite{Hinton2015DistillingTK,Zagoruyko2017PayingMA,Ahn2019VariationalID,Mukherjee2019CogniNetCF}. 
The core of KD is that a previously trained large teacher model transfers knowledge to a small student model, so that the student model performs comparably to the teacher model. Moreover, the student model also has lower inference latency due to its small size and compact network architecture. Feature Distillation (FD) is a primary method in KD, which distill features in the teacher network through a transformation~\cite{Romero2015FitNetsHF, Ahn2019VariationalID, Tian2020ContrastiveRD}. 

Recently, inspired by the success of KD in various fields, it has been introduced into the Recommendation System (RS) to enhance the capability of modeling users' preferences or reduce inference latency while maintaining performance~\cite{Lee2019CollaborativeDF,Tang2018RankingDL,Kang2020DE-RRD}, such as Ranking Distillation (RD), Collaborative Distillation (CD). Ranking Distillation (RD) learns to rank documents/items from both the training data and the supervision of a larger teacher model~\cite{Tang2018RankingDL,Kang2020DE-RRD}. For example, Kang et al.~\cite{Kang2020DE-RRD} proposed a method that enables the student model to learn from the latent knowledge encoded in the teacher model and from the teacher's predictions results. Collaborative Distillation (CD) first samples unobserved items from the teacher's recommendation list according to their ranking, then trains the student to mimic the teacher's prediction score (e.g., relevance probability) on the sampled items~\cite{Lee2019CollaborativeDF}. 

In this paper, we propose a teacher model which extracts multimodal features from the generic feature, guided by semantic information. Meanwhile, we employ a student model with shallow neural networks to serve as feature extractors for various modalities in recommendation model. 

\section {Our Method}
\label{sec:method}

\begin{figure*}[t]
	\centering
	\includegraphics[width=0.8\linewidth]{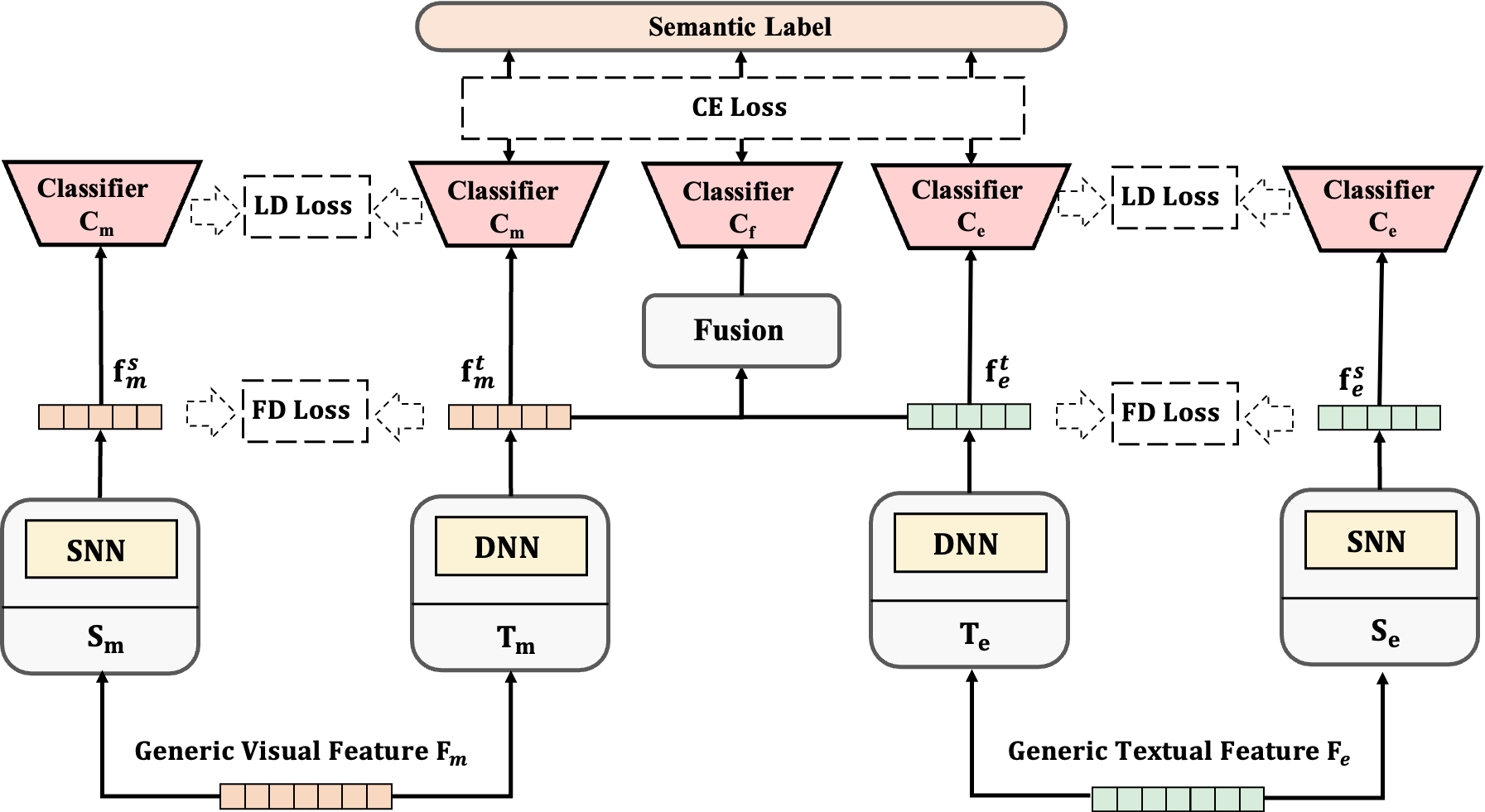}
	\vspace{0.0cm}
	\caption{An overview of the proposed Semantic-Guided Feature Distillation (SGFD) method. Taking visual feature extraction as an example, the visual feature extractor in the teacher model extracts features from generic visual features considering both the semantic and complementary information. The Logits-based Distillation (LD) loss and the Feature-based Distillation (FD) loss are employed to transfer the knowledge to the student model.}
	\label{fig:framework}
\end{figure*}

\subsection{Preliminaries}
\subsubsection{\textbf{Task Formulation.}} 
In this paper, we focus on feature extraction in multimodal recommendation. 
To alleviate the data sparsity problem in recommendation, multimodal features (e.g., user review and item image~\footnote{Note that only the multimodal data of items is used in this work. Besides, each item is associated with one image.}) are leveraged to learn better representations of users and items. In multimodal recommendation, the generic review and image features are extracted from multimodal data using pre-trained models. The main objective of our work is to derive effective recommendation-oriented features from these generic multimodal features to improve recommendation accuracy. In other words, our task is to effectively utilize the information contained in the reviews and images of items to better capture users' preferences and make more accurate recommendations.


\subsubsection{\textbf{Generic Feature Extraction.}} 
In our implementation, we use the VGG16 model~\cite{Simonyan2015VeryDC} to extract generic visual features from item images. This model consists of 13 convolutional layers and three fully connected layers. It is trained by using 1.2 million ImageNet (ILSVRC2010) images. We utilize the output from the second fully connected layer, a 4096-dimensional feature vector, as the generic visual feature $\mathbf{F}_m$. Additionally, we employ BERT~\cite{Devlin2019BERTPO} to extract generic textual features. This model is capable of deriving deep bidirectional representations from the unlabeled text by joint conditioning on both left and right contexts across all layers. In our work, we input all the users' reviews for an item to BERT to obtain the generic textual feature $\mathbf{F}_e$ for this item. 

\subsubsection{\textbf{Recommendation-oriented Feature Extraction.}} 
After obtaining the generic visual and textual features $\mathbf{F}_m$ and $\mathbf{F}_e$, the feature extractors in the multimodal recommendation model are then used to tailor these features for recommendation. Take visual feature as an example, previous studies~\cite{Wei2019GRCN,liu2018MAML,BM32020Arxiv} feed the generic visual feature $\mathbf{F}_m$ into a Multi-Layer Perceptrons (MLP) to extract the recommendation-oriented visual feature $\mathbf{f}_m$. Specifically, it is defined as:

\begin{equation}
\label{upm:feature_map}
    \mathbf{f}_m = {\rm \text{MLP}}(\mathbf{F}_m),
\end{equation}
It should be noted that the adopted ${\rm MLP}(\cdot)$ are jointly optimized with the recommendation model. Thus, the training of feature extractors will also suffer from the data sparsity problem in recommendation. The recommendation-oriented textual feature $\mathbf{f}_s$ can be obtained similarly. In multimodal recommendation models, the textual feature $\mathbf{F}_e$ and visual feature $\mathbf{F}_m$ will be fed into two separate MLPs to extract the effective recommendation-oriented features $\mathbf{f}_e$ and $\mathbf{f}_m$, respectively. These extracted features $\mathbf{f}_e$ and $\mathbf{f}_m$ are subsequently used to improve recommendation accuracy.

\subsection{\textbf{Semantic-guided Feature Distillation}}
We propose a model-agnostic approach, termed semantic-guided feature distillation, to extract effective recommendation-oriented features from generic multimodal features for recommendation. As shown in Fig.~\ref{fig:framework}, the proposed method consists of a teacher model and a student model. In the teacher model, the DNN-based feature extractors ($\text{T}_e$ and $\text{T}_m$) for textual and visual features are designed to extract effective features considering the semantic and complementary information. In our cases, the feature extractor in the student model replaces the original extractor in multimodal recommendation models. In other words, the SNN-based feature extractors ($\text{S}_e$ and $\text{S}_m$) in the student model are part of multimodal recommendation models. Therefore, it can obtain the semantic information from teacher extractors by the adopted knowledge distillation technique and benefit from the user behaviors. Next, we detail these two models and the knowledge distillation technique.
\subsubsection{\textbf{Teacher Model}}
Since semantic information is often overlooked while extracting features from multimodal data for recommendation, we propose a teacher model that extracts effective features considering the semantic information of items. Specifically, we leverage semantic labels (e.g., category labels) as supervised information to classify the extracted features using classifiers. Our proposed method employs three distinct classifiers $\text{C}_e(\cdot)$, $\text{C}_m(\cdot)$ and $\text{C}_f(\cdot)$ for textual, visual, and fused features.
Moreover, considering that the generic modality feature is highly complex and high dimensional, we utilize deep neural networks as feature extractors for visual and textual features, denoted as $\text{T}_m(\cdot)$ and $\text{T}_e(\cdot)$, respectively. 

\textbf{Extracting Semantic-aware Features.}
Take the generic visual feature $\mathbf{F}_m\in \mathcal{R}^{D_m}$ as an example, we first feed $\mathbf{F}_m$ into the feature extractor to obtain valuable features $\mathbf{f}_m^t \in \mathcal{R}^{d_m}$:
\begin{equation}
\label{upm:feature_multi-layer}
    \mathbf{f}_m^t = \text{T}_m(\mathbf{F}_m),
\end{equation}
where $\text{T}_m$ is the feature extractor that consists of a deep neural network with $k$ layers~\footnote{$k$ is calculated by $k=\log_{\delta}{\frac{D_m}{d_m}}$, where $\delta$ is a hyperparameter that controls the number of layers based on the input dimension of each modality.}. Similarly, we obtain the textual feature $\mathbf{f}_e^t$ from the generic textual feature using the textual feature extractor $\text{T}_e$.
The extracted visual and textual features are then separately fed into the classifier $\text{C}_m(\cdot)$ and $\text{C}_e(\cdot)$. Take the visual feature $\mathbf{f}_m^t$ as an example, it is formulated as follows:
\begin{equation}
\label{upm:teacher_extractor_classification}
    \mathbf{P}_m^t = \text{C}_m(\mathbf{f}_m^t;\tau) = \text{softmax}((\mathbf{W}_{m}\mathbf{f}_m^t + \mathbf{b}_m)/ \tau),
\end{equation}
where $\mathbf{P}_m^t$ represents the probability distribution output for the visual feature. $\mathbf{W}_{m}$ and $\mathbf{b}_m$ are the trainable weight matrices and bias vectors, respectively. It should be noted that the classifier parameters for each modality are trained independently. Here, $\tau$ refers to a temperature hyperparameter that controls the smoothness of the distribution. Similarly, the probability distribution output for the textual feature $\mathbf{P}_e^t$ can be obtained.

The feature extractors are optimized using cross-entropy (CE) loss. The combined CE loss for textual and visual features is formulated as follows:


\begin{equation}
\label{upm:techer_extractor_ce_loss}
     \mathcal{L}_{TCE} =  (\text{CE}(\mathbf{P}_m^t, \mathbf{y}) + \text{CE}(\mathbf{P}_e^t, \mathbf{y}))/2,
\end{equation}
where $\text{CE}(\mathbf{\hat{y}}, \mathbf{y}) = - \sum_{i=1}^{C}y_{i} \cdot log(\hat{y}_{i})$ denotes the cross-entropy loss. $C$ is the number of classes, $\mathbf{y}$ is the ground-truth label, and $\mathbf{\hat{y}}$ is the predicted probability distribution.

\textbf{Exploiting Complementary Information.}
To enhance the ability of teacher extractors to extract effective features of each modality, we first leverage complementary information between multiple modalities by fusing textual and visual features. As shown in Fig.~\ref{fig:framework}, taking the output of each modality extractor as input, we fuse the vectors as follows:
\begin{equation}
\label{upm:fusion_loss1}
    \mathbf{f}_u = \sigma(\mathbf{W}_{u}[\mathbf{f}_m^t;\mathbf{f}_e^t]) + \mathbf{b}_{u}),
\end{equation}
where $\mathbf{W}_{u}$ and $\mathbf{b}_{u}$ denote the trainable weight matrix and bias vector. $\sigma(\cdot)$ is the activation function and ReLU is adopted because of its biologically plausible and non-saturated property~\cite{Xavier2010xavier}. Then, the fused feature $\mathbf{f}_u$ is fed into the classifier $\text{C}_f(\cdot)$, and the CE loss function of the fused feature is defined as follows:
\begin{equation}
\label{upm:fusion_loss3}
    \mathbf{P}_u = \text{C}_f(\mathbf{f}_u;\tau),
\end{equation}
\begin{equation}
\label{upm:fusion_loss3}
    \mathcal{L}_{FCE} =  \text{CE} (\mathbf{P}_u, \mathbf{y}).
\end{equation}
The weight matrix and the bias vector of $\text{C}_f(\cdot)$ are trainable and specifically trained with fused features to capture the complementary information between multiple modalities. The temperature hyperparameter $\tau$ is adopted similarly as they are in Eq.~\ref{upm:techer_extractor_ce_loss}. 

\subsubsection{\textbf{Student Model}}
To reduce the negative effect of the data sparsity problem on feature extraction in recommendation, we utilize shallow neural networks (SNNs) as feature extractors for generic multimodal features. The feature extractors for visual and textual features in the student model are denoted as $\text{S}_m(\cdot)$ and $\text{S}_e(\cdot)$, respectively. For the generic visual feature $\mathbf{F}_m$, the recommendation-oriented visual feature $\mathbf{f}_m^s$ is obtained by:

\begin{equation}
\label{upm:fusion_loss3}
    \mathbf{f}_m^s = \text{S}_m(\mathbf{F}_m).
\end{equation}

The textual feature $\bm{f_e^s}$ can be obtained in the same manner. 
Then, the visual and textual features $\mathbf{f}_m^s$, $\mathbf{f}_e^s$ are fed into the classifier $\text{C}_m(\cdot)$ and $\text{C}_e(\cdot)$ respectively. Taking the visual feature as an example, it is formulated as:
\begin{equation}
\label{upm:student_extractor_classification}
    \mathbf{P}_m^s = \text{C}_m(\mathbf{f}_m^s;\tau).
\end{equation}
By following this way, we can obtain $\bm{P}_e^s$ for the textual feature. Note that the classifiers adopted are the same as Eq.\ref{upm:teacher_extractor_classification}. 



\subsubsection{\textbf{Knowledge Distillation}}
The proposed teacher model, which utilizes deep neural networks, is capable of extracting semantic-aware features from the generic features of each modality. On the other hand, the student model, employing SNNs, can alleviate the negative effect of the data sparsity problem on feature extraction in recommendation. To leverage the strengths of both the teacher and student models, we employ the knowledge distillation technique to transfer the response-  and feature-based knowledge encoded in the teacher model to the student model. Next, we detail the two distillation losses used for transferring this knowledge.



\textbf{Response-Based Knowledge.} Response-based knowledge refers to the neural response (logits) generated by the final output layer of the teacher model. In our method, we consider the output of the classifiers as response-based knowledge. Hence, we employ the Logits-based Distillation (LD) loss to enable the feature extractors in the student model to mimic the predicted label distribution of the feature extractors in the teacher model. The distillation loss between the feature extractors of visual and textual features for both the teacher and student models can be described as follows:

\begin{equation}
\label{upm:extract_logits_loss}
    \mathcal{L}_{LD} =   || \mathbf{P}_m^t - \mathbf{P}_m^s||_2^2 + || \mathbf{P}_e^t - \mathbf{P}_e^s||_2^2.
\end{equation}
Note that we do not force the student model to match the prediction distribution of the teacher model; instead, we expect the student model to gain semantic information via knowledge transfer. Thus, following the previous work~\cite{Hinton2015DistillingTK}, we use soft targets to embody the informative knowledge derived from the teacher model. 

\textbf{Feature-Based Knowledge.} Given the capacity of deep neural networks for feature extraction, the output from intermediate layers can serve as knowledge to guide the training of the student model. In our method, we employ the Feature-based Distillation (FD) loss to transfer the visual and textual feature-based knowledge from the teacher model to the student model. The distillation loss can be formulated as follows: 


\begin{equation}
\label{upm:logits_loss}
    \mathcal{L}_{FD} =  || \mathbf{f}_m^t - \mathbf{f}_m^s ||_2^2 + || \mathbf{f}_e^t - \mathbf{f}_e^s ||_2^2.
\end{equation}




\subsection{\textbf{Optimization}}
In this work, we target the top-$n$ recommendation and equip our proposed SGFD with three multimodal recommendation models (MAML~\cite{liu2018MAML}, GRCN~\cite{Wei2019GRCN}, BM3~\cite{BM32020Arxiv}). However, these three recommendation models have different objective functions. Therefore, we refer to the loss of objective functions of these three recommendation models as $\mathcal{L}_{REC}$. 

For our proposed SGFD, the objective function consists of the Cross-Entropy Loss in the teacher model and the Knowledge Distillation Loss. Specifically, the total Cross-Entropy Loss is formulated as:
\begin{equation}
    \mathcal{L}_{CE} =  \mathcal{L}_{TCE} +  \mathcal{L}_{FCE},
\end{equation}
The Knowledge Distillation Loss utilized for knowledge transfer is formulated as:
\begin{equation}
\label{upm:KD_loss}
    \mathcal{L}_{KD} =   \mathcal{L}_{LD} +  \mathcal{L}_{FD},
\end{equation}

The student model of the proposed SGFD is jointly trained with the multimodal recommendation model in an end-to-end manner. Overall, the final loss function of the recommendation model equipped with SGFD is,
\begin{equation}
\label{upm:SemMFD_loss}
   \mathcal{L}_{SGFD} = \mathcal{L}_{REC} +  \lambda_1 \mathcal{L}_{KD} + \lambda_2 \mathcal{L}_{CE},
\end{equation}
where $\lambda_1$ and $\lambda_2$ are used to adjust the weight of the knowledge distillation loss and cross-entropy loss.

\section{Experiments}
\label{sec:experiments}







\subsection{Experimental Setup}
\subsubsection{\textbf{Datasets}}
\begin{table}[t]
	\centering
	\caption{ Basic statistics of the experimental datasets.}
	\label{tab:data}
	\begin{tabular}{cccccl}
		\toprule
		Dataset&\#user&\#item&\#label & \#interactions&sparsity \\
		\midrule
		Office & 4,874 & 7,279 &54 & 52,957 &99.85\% \\
		Clothing & 18,209& 17,318 &26 & 150,889 &99.98\% \\
		Toys Games & 18,748& 11,672& 19 & 161,653 &99.97\% \\ 
		\bottomrule
	\end{tabular}
	\vspace{-0.2cm}
\end{table}

In experiments, we utilize the Amazon review dataset~\footnote{http://jmcauley.ucsd.edu/data/amazon.}  for experimental evaluation as in previous studies~\cite{mcauley2013hidden,liu2018MAML}. 
This public dataset contains user interactions (review, rating, helpfulness votes, etc.) on items and the item metadata (descriptions, attributes, images, etc.) on 24 product categories. To verify the effectiveness of our proposed method, we choose the three per-category datasets (\emph{Office}, \emph{Clothing} and \emph{Toys\&Games}) for performance evaluation under different sizes and sparsity levels. In these datasets, we treat each rating as a positive user interaction. 
In contrast, we treat all the items the user did not interact with as negative. And then, we pre-processed the dataset with a 5-core setting on both items and users in our settings. 
All user reviews and item images are used as item-side information. 
Table~\ref{tab:data} shows the basic statistics of the four datasets.

In this work, we focus on recommending a set of top-$N$ ranked items that will appeal to the user. Specifically, we randomly sampled 80\% of the interactions of each user to create the training set and retained 20\% of the user interactions for testing.

\subsubsection{\textbf{Baselines}}
To demonstrate the effectiveness of our proposed SGFD, we first equipped it on three existing multimodal recommendation models: MAML~\cite{liu2018MAML}, GRCN~\cite{Wei2019GRCN} and BM3~\cite{BM32020Arxiv}. Then, we compared the performance of newly built models (SGFD$_{MAML}$, SGFD$_{GRCN}$, SGFD$_{BM3}$) with the following baselines, including general CF recommendation models (e.g., NeuMF~\cite{he2017neural}, CML~\cite{hsieh2017cml}, NGCF~\cite{wang2019ngcf} and DGCF~\cite{wang2020DGCF}) and multimodal recommendation models (e.g., MAML, GRCN, BM3).

\subsubsection{\textbf{Evaluation Metrics}}
Two widely-used evaluation metrics are adopted for top-$n$ recommendation: \emph{Recall} and \emph{Normalized Discounted Cumulative Gain} (NDCG)~\cite{he2015trirank}. 
Recommendation accuracy is calculated for each metric based on the top 20 results. Note that the reported results are the average values across all testing users. 

\subsubsection{\textbf{Parameter Settings}}
The PyTorch framework~\footnote{https://pytorch.org.} is adopted to implement the proposed model. In our experiments, all hyperparameters are carefully tuned. For all datasets, the embedding size of the users and items is set to 128. The mini-batch size is fixed to 1024. The learning rate for the optimizer is searched from $\{10^{-5},10^{-4},\cdots,10^{+1}\}$, and the model weight decay is searched in the range $\{10^{-5}, 10^{-4},\cdots, 10^{-1}\}$. For knowledge transfer, the temperature $\tau$ is tuned
from$\{5,10,20,50,100\}$, while for classification, it is set to 1. The layer number of the neural networks in the teacher model $\delta$ is tuned from $\{2,4,6,8\}$. 
Moreover, the weights $\lambda_1$ and $\lambda_2$ are searched in the range $\{10^{-3}, 10^{-2},\cdots, 10^{+1}\}$.
Besides, the early stopping strategy is adopted.
Specifically, the training process will stop if recall@20 does not increase for 20 successive epochs. 
\subsection{Study of Multimodal Feature Extraction in Recommendation}
\begin{figure}[t]
\centering
	\subfloat[Office]{\includegraphics[width=0.8\linewidth]{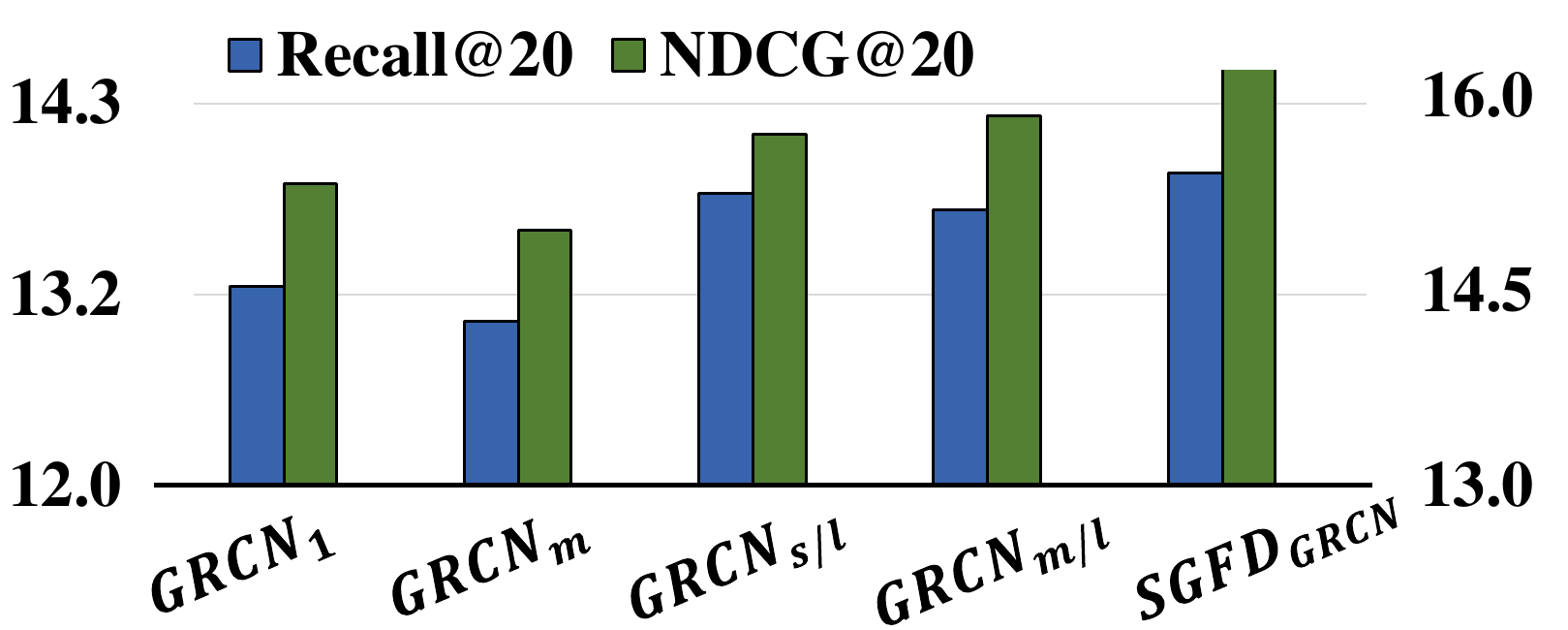}}
        \newline	
	\subfloat[Clothing]{\includegraphics[width=0.8\linewidth]{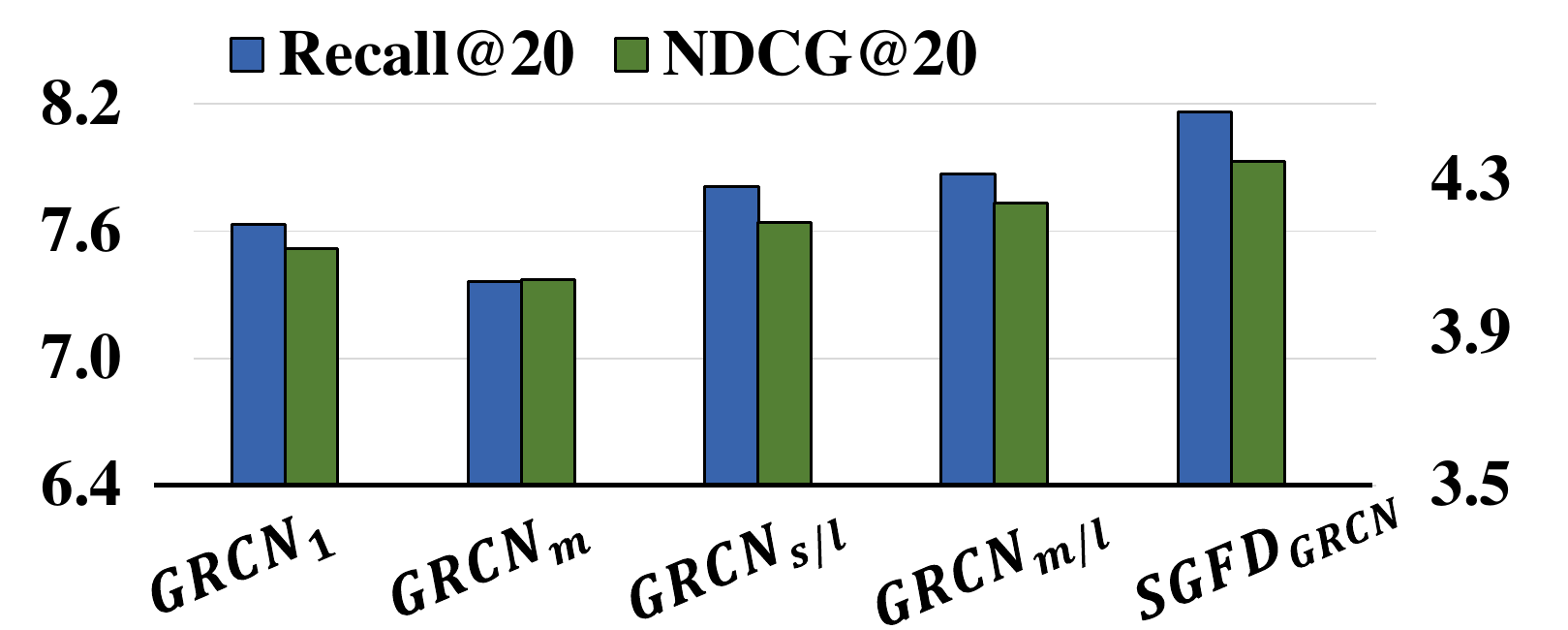}}
        \newline
	
	\vspace{-0.2cm}
	\caption{Performance comparison of GRCN equipped with different feature extraction methods. Notice that the values are reported by percentage with '\%' omitted.}
	\label{fig:similarity}
\end{figure}
In this section, we primarily investigate the influence of different feature extractors for multimodal features on recommendation performance. We compare the performance of four feature extraction methods with our proposed SGFD based on GRCN, and the results are reported in Fig.~\ref{fig:similarity}. Our analyses are based on performance comparisons to the following methods:
\begin{itemize}[align=left,style=nextline,leftmargin=*,labelsep=\parindent,font=\normalfont]
\item \textbf{GRCN$_{1}$~\cite{Wei2019GRCN}}: It is the vanilla GRCN method that adopts the shallow neural network as the feature extractor.
\item \textbf{GRCN$_{m}$}: The vanilla GRCN method adopts the deep neural network as the feature extractor.
\item \textbf{GRCN$_{s/l}$}: Given the vanilla GRCN method, we employ the teacher model of SGFD with shallow neural networks as the feature extractor in GRCN. 
\item \textbf{GRCN$_{m/l}$}: Given the vanilla GRCN method, we employ the teacher model of SGFD with the deep neural networks as the feature extractor in GRCN.
\end{itemize}


From the results, we can see that GRCN$_{1}$ yields better performance than GRCN$_{m}$ on \emph{Office} and \emph{Clothing}. Furthermore, the performance degradation of GRCN$_{m}$ in the \emph{Office} is less compared to \emph{Clothing}. This can be attributed to the fact that the feature extractor in most multimodal recommendation models is jointly optimized with recommendation models. Consequently, the feature extractor will suffer the data sparsity problem when the user-item interactions are sparsity. As \emph{Office} is denser, the data sparsity problem less impacts the feature extraction model. 
The variant GRCN$_{s/l}$ outperforms GRCN$_1$ and GRCN$_m$ across all the datasets, which indicates that the semantic information has positive advancement in extracting valuable features for recommendation.
More importantly, by adopting multiple neural networks and introducing the complementary information of multimodal features, the variant GRCN$_{m/l}$ has an explainable improvement on \emph{Clothing}. This demonstrates that the complementary information is facilitated to reduce redundant information derived from the feature extractor in each modality. But, it still underperforms GRCN$_{s/l}$ on \emph{Office} in terms of Recall. The reason might be that it also suffers from the data sparsity problem when jointly training the extractor with recommendation models. The SGFD$_{GRCN}$ outperforms all variants over all datasets, which demonstrates the effectiveness of our proposed SGFD.

\subsection{\textbf{Performance Comparison}}
\label{sec:experiments_pc}
\begin{table}[t]
	\caption{Performance of our SGFD model and the competitors over three datasets. Notice that the values are reported by percentage with '\%' omitted.} 
	\centering
 \resizebox{\linewidth}{!}{
		\begin{tabular}{l|cc|cc|cc}
		\hline
    Datasets   & \multicolumn{2}{c|}{Office} & \multicolumn{2}{c|}{Clothing} & \multicolumn{2}{c}{Toys\&Games} \\ \hline
    Metrics    & Recall         & NDCG    & Recall          & NDCG    & Recall         & NDCG        \\ \hline\hline
    NeuMF      & 6.05       & 3.37    & 1.89          & 0.80    & 2.53           & 1.28        \\
    CML        & 12.29      & 13.95   & 4.09          & 2.24    & 12.27         & 11.60       \\
    NGCF       & 8.14       & 4.79    & 4.66          & 2.16    & 9.70           & 5.87        \\ 
    DGCF       & 12.06      & 11.05   & 7.45          & 4.05    & 12.62         & 10.85       \\ \hline\hline
    MAML       & 13.33      & 14.12   & 6.89          & 3.70    & 12.84          & 12.09       \\
    SGFD$_{MAML}$ & \textbf{14.22*}      & \textbf{14.87*}   & \textbf{7.18*}         & \textbf{3.89*}    & \textbf{13.23*}          & \textbf{12.60*}      \\ 
    Improv.    & 6.68\%    & 5.31\%  & 4.21\%    & 5.14\%  & 3.04\%        & 4.22\%   \\ \hline\hline
    GRCN       & 13.20      & 15.33   & 7.63     & 4.12    & 11.85          & 11.23       \\
    SGFD$_{GRCN}$ & \textbf{13.88*}      & \textbf{16.26*}   & \textbf{8.16*}        & \textbf{4.35*}    & \textbf{12.59*}          & \textbf{11.93*}     \\ 
    Improv.    & 5.15\%    & 6.07\%  & 6.95\%     & 5.58\%  & 6.24\%        & 6.23\%    \\ \hline\hline
    BM3        & 13.67      & 15.36    & 7.43      & 4.21    & 12.22          & 11.92        \\  
    SGFD$_{BM3}$  & \textbf{14.07*}     & \textbf{15.86*}    & \textbf{7.84*}      & \textbf{4.57*}    & \textbf{12.54*}          & \textbf{12.94*}       \\ 
    Improv.    & 2.93\%    & 3.25\%  & 5.52\%    & 8.56\%  & 2.62\%       & 8.56\%   \\ \hline
		\end{tabular}}
		\begin{tablenotes}
		\item The symbol * denotes that the improvement is significant with $p-value < 0.05$ based on a two-tailed paired t-test.
	\end{tablenotes}
		\label{tab:results}
		\vspace{-0.2cm}
\end{table}

We report the performance of three multimodal recommendation models equipped with our proposed SGFD and all the competitors over the three datasets in table~\ref{tab:results}. Our proposed approach achieves significant performance improvements over all the competitors across all the datasets in terms of different metrics. From the experiment results, we achieve some compelling observations.

Among all competitors, NeuMF, CML, NGCF and DGCF are solely trained using user-item interactions. NeuMF achieves promising performance in comparison to traditional MF-based methods~\cite{he2017neural}, due to the capability of deep neural networks to model the non-linear interactions between users and items. 
The dot product cannot well model fine-grained user preferences, as it does not obey the triangle inequality. Therefore, CML replaces the dot product with metric distance, improving performance compared to NeuMF. Benefiting from the high-order information, NGCF accomplishes state-of-the-art results. 
Moreover, DGCF achieves better performance than NGCF by leveraging the disentangled representation technique to learn robust and independent user and item embeddings.  
MAML, GRCN, and BM$3$, which utilize multimodal features, generally achieve better performance than those methods relying solely on user-item interactions.
Specifically, MAML outperforms CML by modeling users' diverse preferences by leveraging multimodal features of items. GRCN surpasses NGCF by employing modality features to discover and prune potential false-positive edges on the user-item interaction graph. BM$3$ consistently performs better than all general CF baselines over all the datasets. It should be credited to the joint optimization of three multimodal objectives to learn the representations of users and items.

To demonstrate the effectiveness of our proposed method, we equip MAML, GRCN and BM$3$ with SGFD, which are denoted as SGFD$_{MAML}$, SGFD$_{GRCN}$ and SGFD$_{BM3}$, by employing the feature extractor in student model to replace the original feature extractor in the recommendation models. 
Note that our approach focuses on distilling knowledge from the teacher to the student model. Hence, the users and items representation learning approaches of the base models (MAML, GRCN and BM$3$) are not modified.
The performance of SGFD$_{MAML}$, SGFD$_{GRCN}$ and SGFD$_{BM3}$ are reported in Table~\ref{tab:results}. From the results, we can see that both SGFD$_{MAML}$, SGFD$_{GRCN}$ and SGFD$_{BM3}$ obtain a significant improvement over all datasets. We credit this to the combined effects of the following three aspects. 
Firstly, the teacher model can extract semantic-aware features from generic multimodal features. Secondly, the knowledge encoded in the teacher model is transferred to the student model by the response- and feature-based knowledge distillation losses. Thirdly, the feature extractors with shallow neural networks in the student model replace the original feature extractors in the base models. Based on this, existing multimodal models can ease the negative impact of the data sparsity problem on representation learning.

\subsection{\textbf{Ablation Study}}
\label{sec:ablation}
\begin{table}[t]
\caption{Performance of our proposed SGFD method equipping on GRCN and their variants over three datasets. The best results are highlighted in bold. Notice that the values are reported by percentage with '\%' omitted.}
\centering
\resizebox{\linewidth}{!}{
\begin{tabular}{l|cl|cl|cc}
\hline
Datasets   & \multicolumn{2}{c|}{Office} & \multicolumn{2}{c|}{Clothing} & \multicolumn{2}{c}{ToysGames} \\ \hline
Metrics    & Recall      & NDCG   & Recall    & NDCG & Recall        & NDCG      \\ \hline
SGFD$_{l}$ & 13.30       & 15.80   & 7.69     & 4.06 & 12.01         & 11.31     \\
SGFD$_{f}$ & 13.53       & 15.71   & 7.86     & 4.18 & 11.55         & 10.92     \\
SGFD$_{e}$ & 13.71       & 15.73   & 7.97     & 4.28 & 12.08         & 11.36     \\
SGFD$_{s}$ & 13.76       & 16.07   & 8.02     & 4.30 & 12.22         & 11.50     \\
SGFD       & \textbf{13.88}     & \textbf{16.26}  & \textbf{8.16}   & \textbf{4.35} & \textbf{12.59}   & \textbf{11.93}   \\ \hline
\end{tabular}}
\label{tab:Ablation}
\vspace{-0.2cm}
\end{table}


In this section, we examine the contribution of different components to the performance of our method, which is equipped with the recommendation method (GRCN). Our analyses are based on performance comparisons with the following variants of our method.
\begin{itemize}[align=left,style=nextline,leftmargin=*,labelsep=\parindent,font=\normalfont]
\item \textbf{SGFD$_{l}$}: It is a variant of SGFD that uses only logits-based distillation loss for knowledge distillation.
\item \textbf{SGFD$_{f}$}: It is a variant of SGFD that uses only feature-based distillation loss for knowledge distillation.
\item \textbf{SGFD$_{e}$}: This variant removes the complementary information exploited from multimodal features in the teacher model.
\item \textbf{SGFD$_{s}$}: This variant adopts the shallow neural network as the feature extractor in the teacher model.
\end{itemize}

From the results, the SGFD$_{l}$ and SGFD$_{f}$ gain different improvements over three datasets, which indicates the benefits from the response-based knowledge and the feature-based knowledge are different. 
Benefiting from using both the logits- and the feature-based distillation loss, SGFD$_{e}$ achieves a remarkable performance across all the datasets. 
SGFD$_{s}$ outperforms SGFD$_{e}$ across all datasets in terms of NDCG. This is because the complementarity between different modality features is taken into account during the feature extraction process.
Our proposed SGFD method achieves superior performance across all datasets. It demonstrates the effectiveness of using the feature extractor with deep neural networks to extract the semantic-aware feature. 

\subsection{\textbf{Visualization of Knowledge Transfer}}

\begin{figure*}[t]
	\centering
	\subfloat[Original Feature Extractor (review)]{\includegraphics[width=0.33\linewidth]{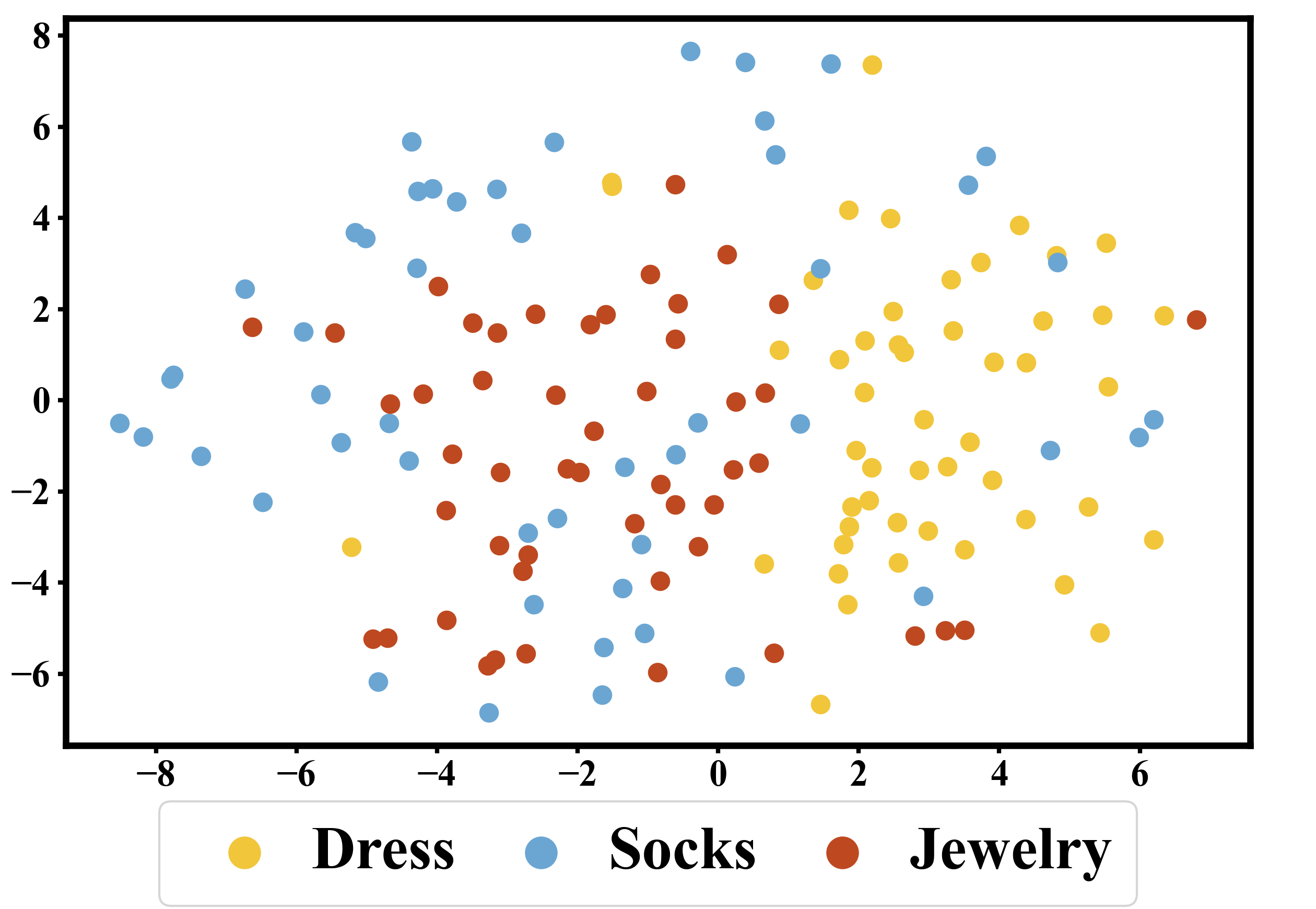}}
	\subfloat[Teacher Model (review)]{\includegraphics[width=0.33\linewidth]{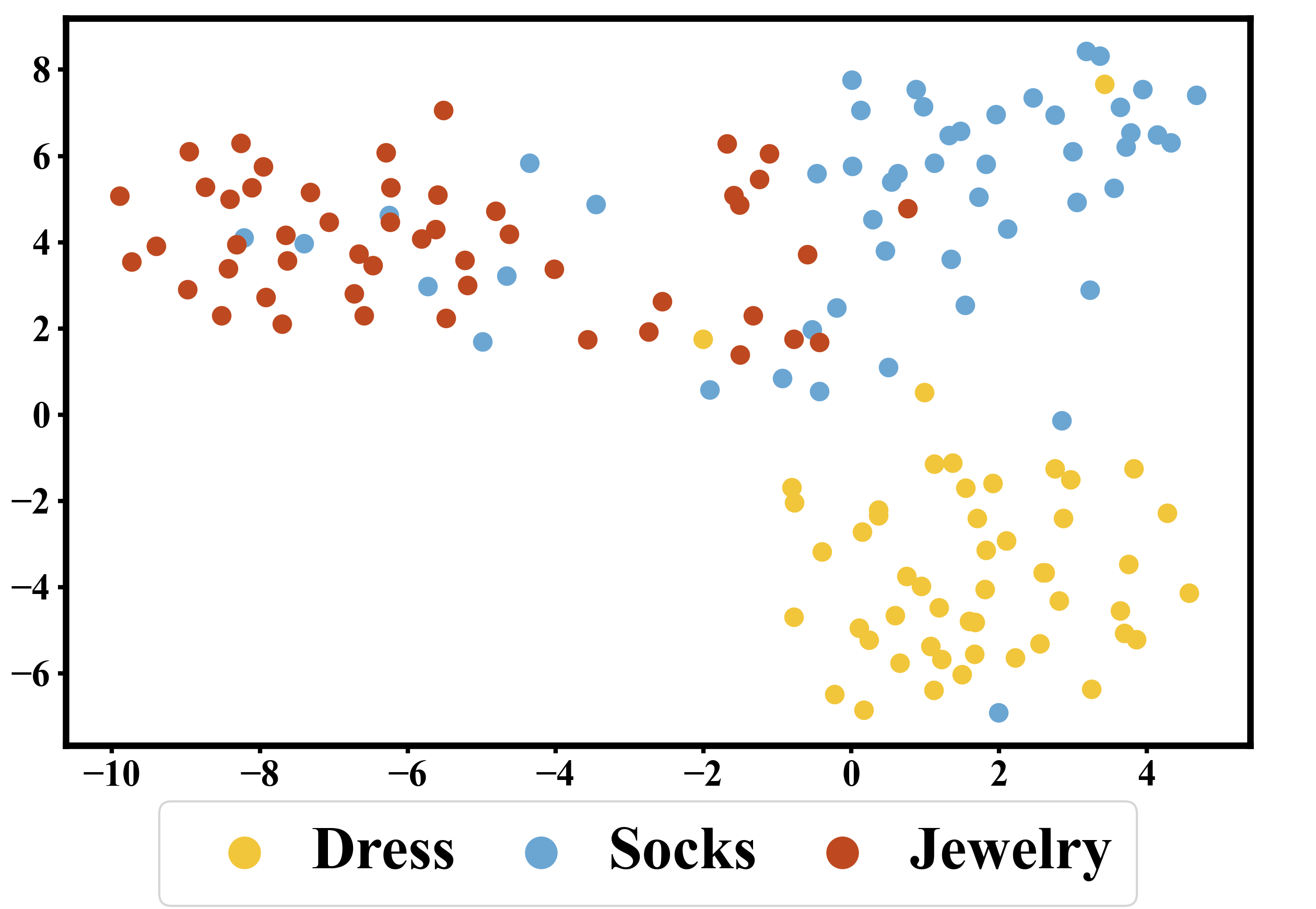}}
	\subfloat[Student Model (review)]{\includegraphics[width=0.33\linewidth]{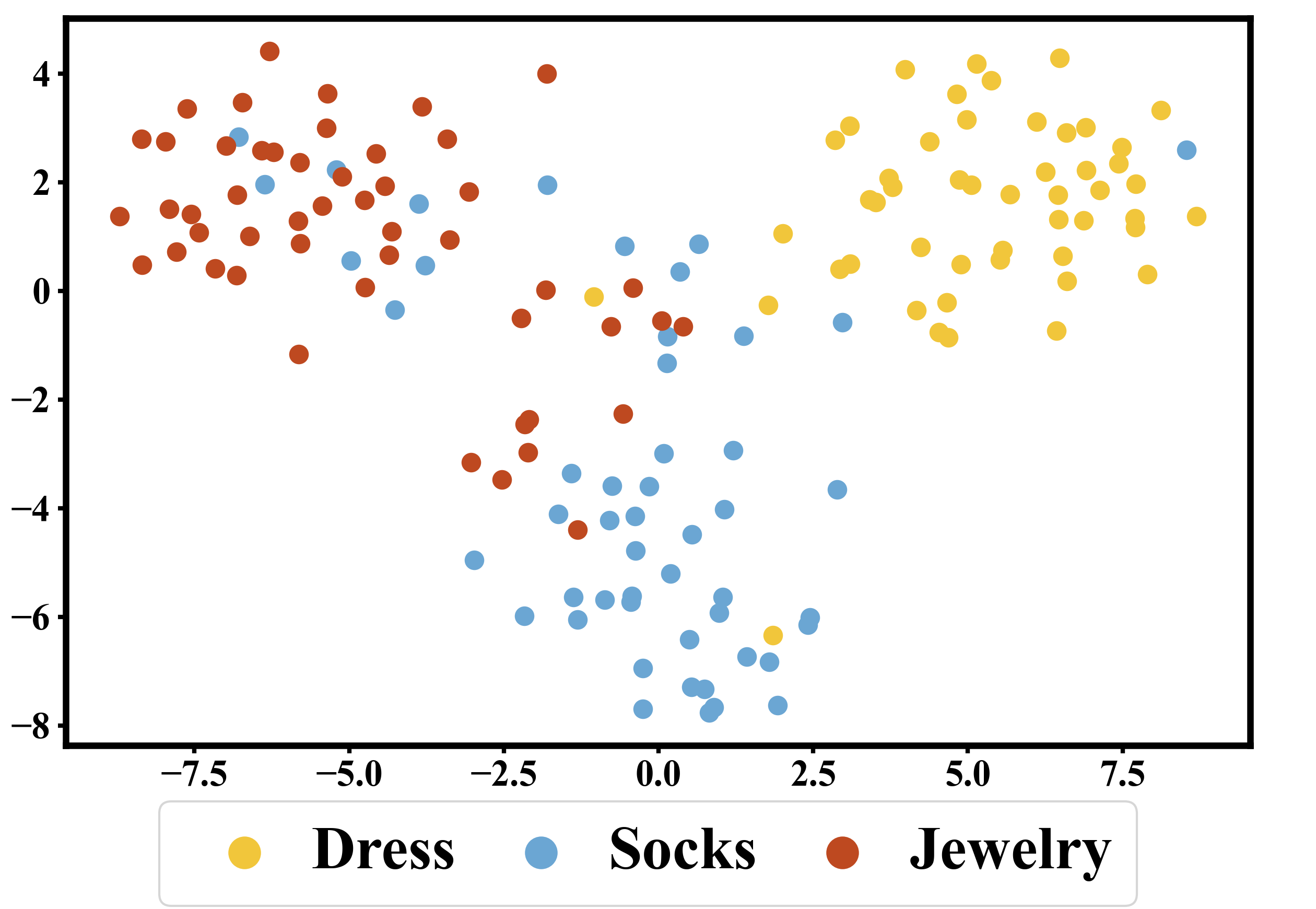}}
	\newline
	\subfloat[Original Feature Extractor (image)]{\includegraphics[width=0.33\linewidth]{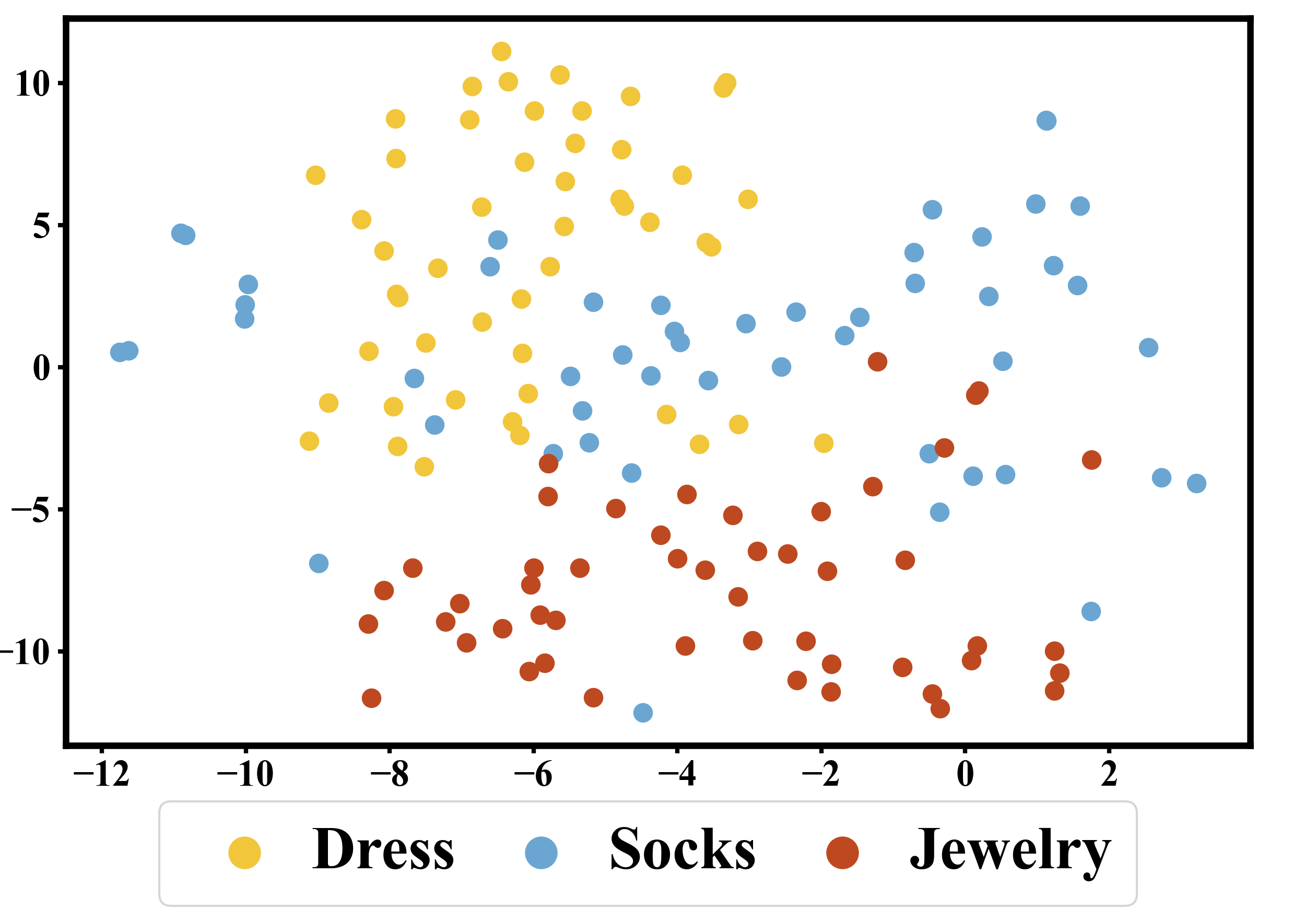}}
	\subfloat[Teacher Model (image)]{\includegraphics[width=0.33\linewidth]{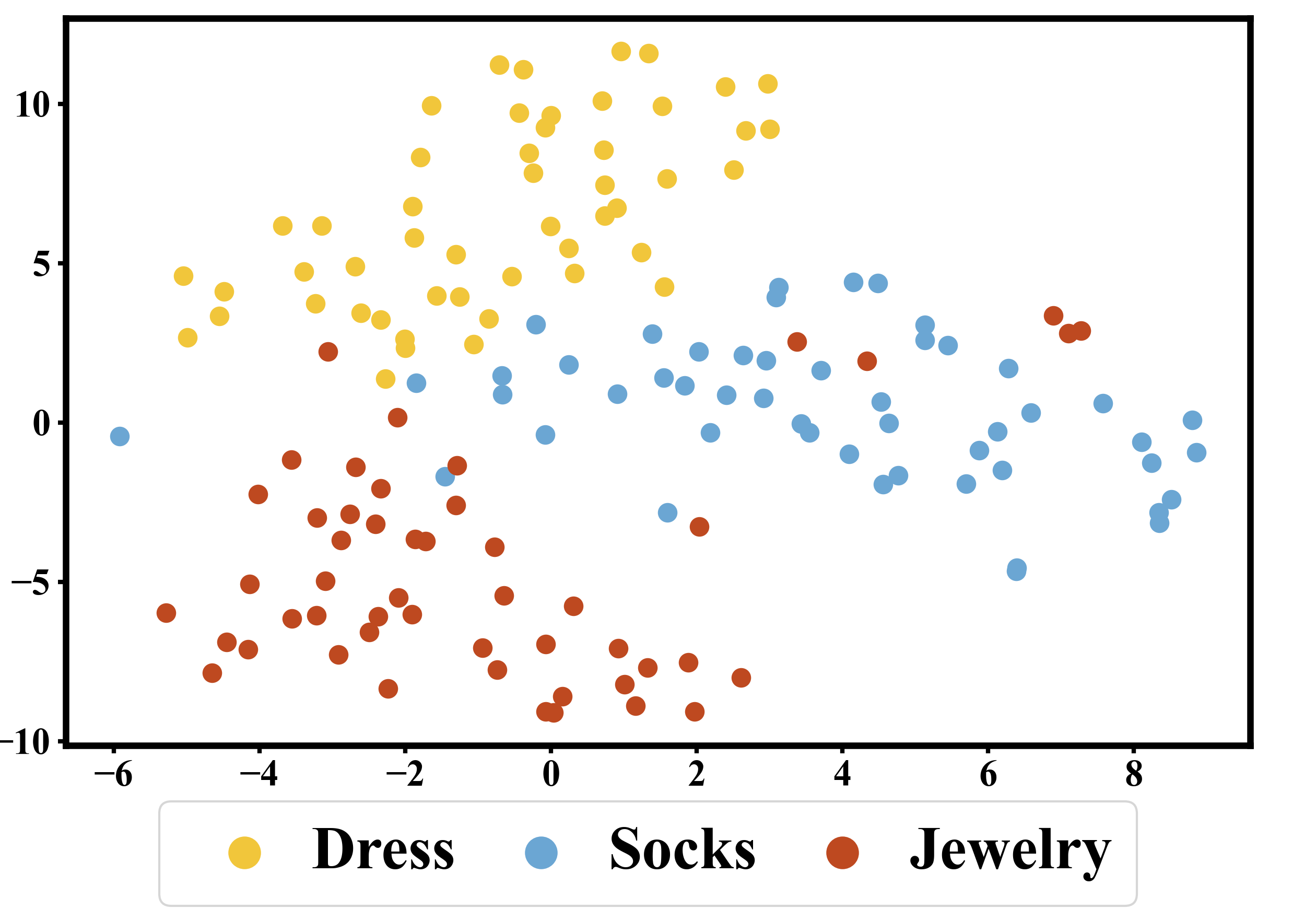}}
	\subfloat[Student Model (image)]{\includegraphics[width=0.33\linewidth]{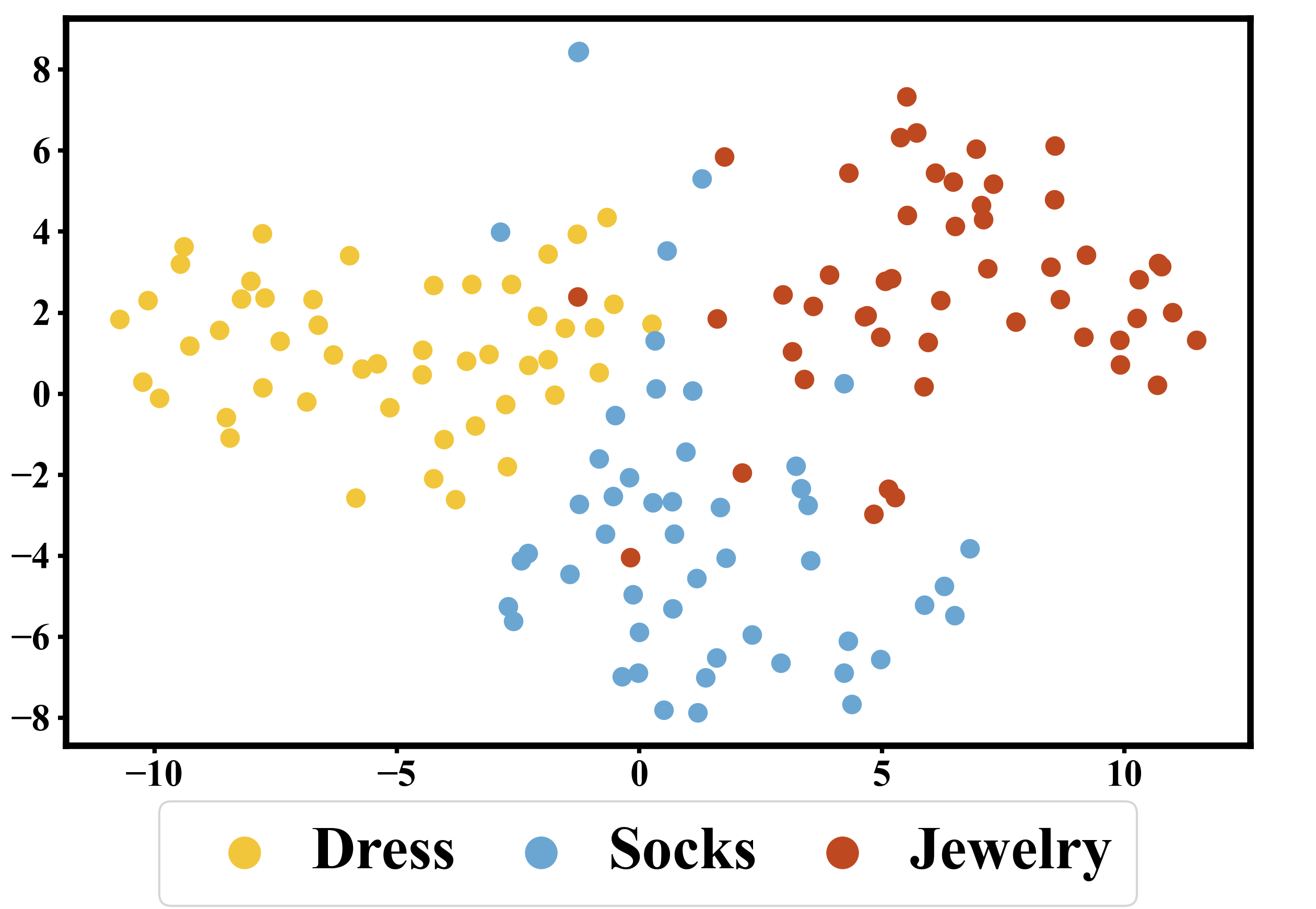}}
	\newline
	\vspace{-0.1cm}
	\caption{Visualization of the vectors extracted from generic review and textual features. The vectors are derived from the original feature extractor in GRCN, feature extractors in teacher and student models of SGFD$_{GRCN}$. In addition, the items are sampled from three categories of \emph{Clothing}: Socks (Orange), Dress (Blue) and Jewelry (Green).}
	\label{fig:category_division}
\end{figure*}
To confirm the transfer of knowledge from the teacher model to the student model, we visualized the review feature and image feature vectors extracted by the original feature extractor in GRCN as well as the feature extractors in the teacher and student model of SGFD$_{GRCN}$, respectively. 

In Fig.~\ref{fig:category_division}, the top three figures show the feature vectors extracted from the generic review feature of items. Meanwhile, the following three figures illustrate the vectors extracted from the generic image feature of items. These items are randomly sampled from three categories of \emph{Clothing} (accessories, body jewelry and socks). Each dot in the figures denotes the extracted vector from the feature of an item. Besides, the dots with the same color in all figures denote the same category of items. The obtained vectors are clustered and visualized via the t-Distributed Stochastic Neighbor Embedding (t-SNE).
As shown in Fig.~\ref{fig:category_division}, for the feature vectors extracted from different extractors, those items of the same category from the teacher extractor are closer than the original extractor, and the boundaries between categories are also more straightforward in Fig.~\ref{fig:category_division}(b). This is attributed to using semantic information to guide the feature extraction process in the teacher model. Besides, the clustering results shown in Fig.~\ref{fig:category_division}(c) demonstrate that the knowledge encoded in the teacher model is transferred to the student model. A similar trend can be observed from the clustering result of feature vectors extracted from generic image features based on the three extractors.

\section{Conclusion}
\label{sec:conclusion}
In this work, we propose a novel model-agnostic approach, named Semantic-Guided Feature Distillation (SGFD), which can robustly extract effective recommendation-oriented features from generic modality features for recommendation. In our approach, feature extractors in the teacher model are proposed to extract effective features considering both the semantic information of items from labels and the complementary information of multiple modalities. Moreover, 
to reduce the negative effect of the data sparsity problem on feature extraction in recommendation, we employ the student model consisting of feature extractors with shallow neural networks. By transferring the knowledge encoded in the teacher model to the student model and replacing the original feature extractor in recommendation methods with the feature extractors in the student model of our SGFD, effective recommendation-oriented features can be obtained. The experimental results on three real-world datasets demonstrate our approach can significantly improve the recommendation accuracy of multimodal recommendation methods.
\section{Acknowledgments}
This research is supported by the National Research Foundation, Singapore under its Strategic Capability Research Centres Funding Initiative, the National Natural Science Foundation of China under Grants 62272254, the Shandong Project towards the Integration of Education and Industry under Grants 2022PY009 and 2022PYI001. Any opinions, findings and conclusions or recommendations expressed in this material are those of the author(s) and do not reflect the views of National Research Foundation, Singapore.

\bibliographystyle{ACM-Reference-Format}
\bibliography{SGMFD}

\end{document}